\documentclass[article]{IEEEtran}
\usepackage{cite}
\usepackage[cmex10]{amsmath}
\usepackage{amssymb}
\usepackage{amsthm}
\usepackage{mathrsfs}
\usepackage{bm}
\usepackage[mathscr]{eucal}
\usepackage{amssymb,amsmath,amsthm,amsfonts,latexsym}
\usepackage{amsmath,graphicx,bm,xcolor,url}
\usepackage{graphicx}
\usepackage{latexsym}
\usepackage{CJK}
\usepackage{indentfirst}
\usepackage{geometry}
\usepackage{psfrag}
\usepackage{setspace}
\usepackage{algorithmic}
\usepackage{algorithmic, cite}
\usepackage{algorithm}
\usepackage{array}
\usepackage{mdwmath}
\usepackage{mdwtab}
\usepackage{eqparbox}
\usepackage{url}
\usepackage{epstopdf}
\usepackage{epsfig,epsf,psfrag}
\usepackage{fixltx2e}
\usepackage{verbatim}
\usepackage{textcomp}
\usepackage{psfrag} 

\usepackage{caption}
\usepackage{subcaption}
\usepackage{tabularx}
\usepackage{graphics}
\usepackage{bbm}

\title{Intelligent User Association for Symbiotic Radio Networks using Deep Reinforcement Learning}

\author{Qianqian Zhang, Ying-Chang Liang, \emph{Fellow, IEEE}, and H. Vincent Poor, \emph{Fellow, IEEE} \\

\thanks{
This work is supported by National Natural Science Foundation of China under Grants 61631005, U1801261, and 61571100. (\emph{Corresponding author: Ying-Chang Liang.})

Q.~Zhang is with the National Key Laboratory of Science and Technology on Communications, and the Center for Intelligent Networking and Communications (CINC), University of Electronic Science and Technology of China (UESTC), Chengdu 611731, China (e-mail: qqzhang kite@163.com).

Y.-C. Liang is with the Center for Intelligent Networking and Communications (CINC), University of Electronic Science and Technology of China (UESTC), Chengdu 611731, China (e-mail: liangyc@ieee.org).

H. V. Poor is with the Department of Electrical Engineering, Princeton University, Princeton, NJ 08544 USA (e-mail: poor@princeton.edu).
}}

\begin{document}

\maketitle

\begin{abstract}

In this paper, we are interested in symbiotic radio networks, in which an \emph{Internet-of-Things} (IoT) network parasitizes in a primary network to achieve spectrum-, energy-, and infrastructure-efficient communications. Specifically, the BS serves multiple cellular users using \emph{time division multiple access} (TDMA) and each IoT device is associated with one cellular user for information transmission. We focus on the user association problem, whose objective is to link each IoT device to an appropriate cellular user by maximizing the sum rate of all IoT devices. However, the difficulty in obtaining the full real-time channel information makes it difficult to design an optimal policy for this problem. To overcome this issue, we propose two \emph{deep reinforcement learning} (DRL) algorithms, both use the historical information to infer the current information in order to make appropriate decisions. One algorithm, centralized DRL, makes decisions for all IoT devices at one time with global information. The other algorithm, distributed DRL, makes a decision only for one IoT device at one time using local information. Finally, simulation results show that the two DRL algorithms achieve comparable performance as the optimal user association policy which requires perfect real-time information, and the distributed DRL algorithm has the advantage of scalability.

\end{abstract}


\section{Introduction}
\label{sec:intro}

The exponential growth in the number of \emph{Internet-of-Things} (IoT) devices will lead to an enormous demand on wireless spectrum and network infrastructure \cite{andrews2014will,Wang2017A,zhang2016spectrum}. To support massive IoT connections, it is highly desirable to design spectrum-, energy-, and infrastructure-efficient communication technologies. \emph{Symbiotic radio networks} (SRN) \cite{zhang2019backscatter,long2019full,guo2019resource}, in which an IoT network parasitizes in a primary network, is envisioned as a promising technique to achieve this goal. In addition, when \emph{ambient backscatter communication} (AmBC) \cite{liu2013ambient} is used for IoT transmission, the IoT devices in SRN transmit their messages to theirs destinations by reflecting the signals received from the primary transmitter without requiring active \emph{radio-frequency} (RF) transmitter chain. That means, the data transmission of the IoT device uses the passive radio technology and does not require dedicated spectrum and infrastructure.
As such, SRN has attracted increasing attention from both academia and industry recently~\cite{wang2016ambient, qian2017semi, ZhangLiangGlobecom17,yang2018modulation,yang2018cooperative, guo2019exploiting,zhang2019constellation,kang2018riding}.

In AmBC-based SRN, the IoT network is an always beneficial party, and thus there are three types of symbiotic relationships based on the interaction between the two coexisting networks: parasitism, commensalism, and mutualism.
Consider a simple SRN model, which consists of three nodes: an RF source, a backscatter IoT device, and a reader. The IoT device backscatters the ambient RF signal by changing its reflection coefficient, through which the information of the IoT device is transmitted to the reader. When the backscatter link is relatively strong as compared to the direct link, and the IoT and primary transmissions have the same baud rate, the IoT gains the transmission opportunity, but it causes severe interference to the primary transmission. Thus, the two networks form the parasitism relationship \cite{zhang2019backscatter,kang2018riding}. When the backscatter link is very weak as compared to the direct link, the effect of backscatter link on the primary transmission is negligible. Thus, in this case, the two networks form the commensalism relationship. Due to the weak backscatter signal, in general, the IoT transmission is much slower than the primary transmission to enhance the transmission performance. Energy detector, which is simple and easy to accomplish, is used in \cite{liu2013ambient,wang2016ambient,qian2017semi,ZhangLiangGlobecom17} to recover the IoT message. However, since the direct link signal is treated as interference in energy detector, the performance suffers from degradation. The performance of the IoT transmission can be improved through interference cancellation \cite{yang2018modulation, guo2019exploiting} or cooperative receiver \cite{yang2018cooperative,zhang2019constellation}.



%

In fact, the backscatter link signal contains the RF source signal and the IoT transmission rate is typically much lower than the primary transmission rate. Thus, the backscatter link can be seen as an additional path of the primary transmission and the slowly changing reflection coefficient of the IoT device introduces time variation for the channel. This observation indicates that the existence of the IoT transmission can improve the performance of the primary system. To achieve it, we need the cooperation between the IoT transmission and the primary network. An example of the cooperation is that the primary receiver and the reader are integrated as a cooperative receiver, which decodes the messages not only from the RF source, but also from the IoT device. In \cite{yang2018cooperative}, the signal detection problem is considered for this scenario and the results show that the existence of the backscatter link benefits the detection of the RF source message based on the joint decoding. Thus, the cooperative design can achieve a mutualism relationship between the primary and IoT transmissions. 

In this paper, we are interested in the user association problem for AmBC-based SRN. The base station (BS) in the primary network serves the cellular users through \emph{time division multiple access} (TDMA), and each IoT device is associated with one cellular user for information transmission by reflecting the signals received from the BS, and each cellular user decodes the messages from the BS and the associated IoT devices using the \emph{successive interference cancelation} (SIC) strategy.
For user association problem in SRN, the BS determines which cellular user an IoT device should be associated with in order to maximize the sum rate of all IoT devices.

In order to obtain the optimal user association strategy, the full real-time channel information is required. However, it is impractical for the BS to obtain all channel information since it involves a great amount of overhead. To overcome this challenge, we use \emph{deep reinforcement learning} (DRL) approach to infer the real-time channel information by using the historical channel knowledge\footnote{When decoding messages, the cellular user needs to estimate the channel information, from which process, the historical channel information is obtained.} based on the channel correlation between different frames. We propose two DRL algorithms, referred to as centralized DRL algorithm and distributed DRL algorithm, to make proper decisions for the user association problem. The centralized DRL algorithm uses the historical global information as the current state to make decisions for all IoT devices at one time, while the distributed DRL algorithm uses the historical local information\footnote{The local information represents the available information at one IoT device.} as the current state to make a decision for one IoT device at one time. Compared with the centralized DRL algorithm, the distributed DRL algorithm has the advantage of scalability, though at the cost of a slightly more information.

In a nutshell, the main contributions of this paper are summarized as follows:
\begin{itemize}
  \item We formulate the user association problem in SRN, which is a challenging task especially for complicated environment.
  \item We propose two DRL-based user association algorithms, namely, centralized DRL and distributed DRL, without the requirement of the full real-time channel information.
  \item The two DRL algorithms use the historical channel information to infer the current information for decision making.
  \item We show that the two proposed DRL algorithms can achieve a performance close to that of the optimal policy with perfect real-time channel information.
  \item Finally, we show that the centralized DRL algorithm needs less information to converge while the distributed DRL algorithm is scalable.
\end{itemize}

\emph{Related Works:}
Recently, DRL has been widely and successfully applied in wireless communication systems, see \cite{luong2018applicationsDeep} for an excellent overview. In particular, in \cite{anh2018deep}, the authors study the time scheduling problem in RF-powered backscatter cognitive radio systems using DRL to maximize the total transmission rate. In \cite{chu2018reinforcement}, a DRL-based access channel control problem is studied for the uplink wireless system with limited access channels. A DRL-based algorithm is proposed in \cite{zhang2018deep} to select proper modulation and coding scheme in cognitive heterogenous networks by learning the interference patten. DRL is adopted in \cite{he2017deep} to schedule users in order to enhance the sum rate in a caching network. In \cite{wang2018handover}, DRL algorithm is used to reduce the handover rate under a constraint of the minimum sum rate. In \cite{yu2018deep}, a distributed DRL multiple access algorithm is proposed to enhance the uplink sum rate in a multi-user wireless system. DRL approach is used in \cite{zhao2018deep} for user association and resource allocation in heterogeneous networks to maximize the overall network utility. Handoff policy in mmWave scenario is studied in \cite{sun2018smart} by taking into account  the mmWave channel characteristics and the \emph{quality of service} (QoS) requirements of users.
Distributed dynamic power control problem is studied in \cite{nasir2018deep} for wireless networks using DRL algorithm.

\emph{Organization:} The rest of the paper is organized as follows. In Section II, the SRN model is established in detail. In Section III, we formulate the user association problem and analyze the optimal policy. Section IV present the two proposed DRL algorithms. Section V presents substantial simulation results for demonstrating the performance. Finally, the paper is concluded in Section VI.

\section{System Model}
\label{sec:system model}

\begin{figure}
\centering
\includegraphics[width=.99\columnwidth] {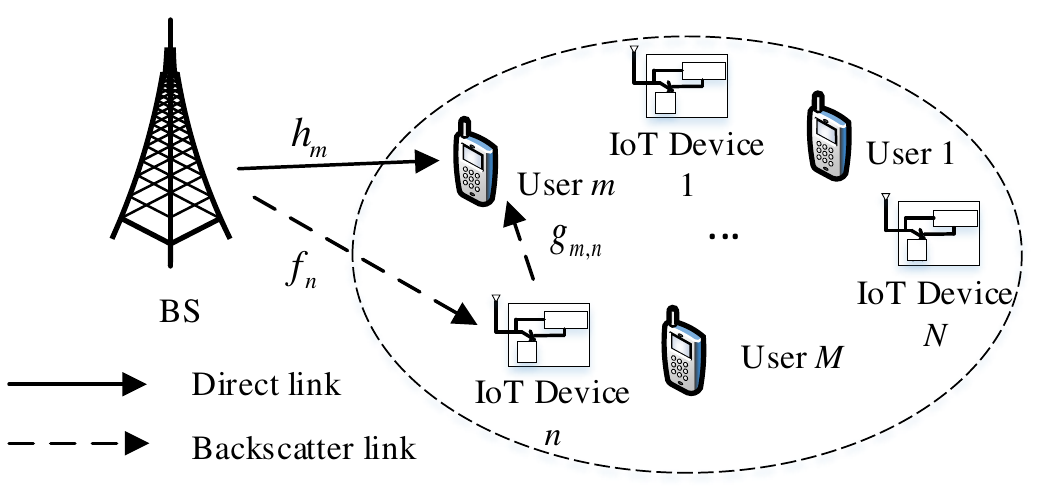}
\captionsetup{font={scriptsize}}
\caption{System model.}
\label{fig:system model}
\vspace{-1.5em}
\end{figure}

\begin{figure}
\centering
\includegraphics[width=.99\columnwidth] {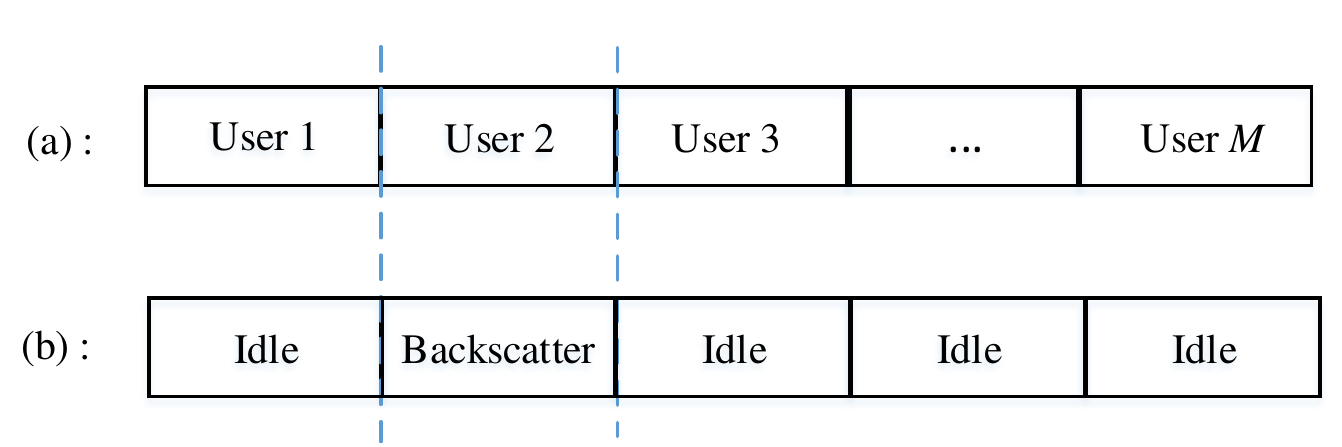}
\captionsetup{font={scriptsize}}
\caption{Frame structure of: (a) the primary network; (b) IoT device $n$.}
\label{fig:frame}
\vspace{-1.5em}
\end{figure}

The system model for the SRN considered in this paper is shown in Fig. \ref{fig:system model}, in which an IoT network parasitizes in a primary network. In particular, the BS in the primary network serves $M$ cellular users through TDMA manner (see Fig. \ref{fig:frame}), while $N$ IoT devices in the IoT network transmit their messages to the associated cellular users by reflecting the received signals from the BS. Specifically, as shown in Fig. \ref{fig:frame}, each IoT device only transmits information in one time slot corresponding to one associated cellular user. The cellular user decodes the signals from both the BS and the associated IoT devices using SIC strategy. In the following, we provide the channel model, the signal model, and the \emph{signal-to-interference-plus-noise ratio} (SINR) model for the SRN.

\subsection{Channel Model}

Here, each channel in the SRN consists of two components: a large-scale fading component and a small-scale fading component.
Denote by $h_m$ the channel coefficient from BS to User $m$ with $h_m = \sqrt{\lambda_{m}}\tilde{h}_m$, by $f_n$ the channel coefficient from BS to IoT Device $n$ with $f_n = \sqrt{\lambda_{n}}\tilde{f}_n$, and by $g_{m,n}$ the channel coefficient from IoT Device $n$ to User $m$ with $g_{m,n} = \sqrt{\lambda_{m,n}}\tilde{g}_{m,n}$, where $\lambda_{m}$, $\lambda_{n}$, and $\lambda_{m,n}$ represent the corresponding large-scale fading components, and $\tilde{h}_m$, $\tilde{f}_n$, and $\tilde{g}_{m,n}$ represent the corresponding small-scale fading components. The large-scale fading components remain unchanged for a fixed distance between the two corresponding nodes, while the small-scale fading components remain unchange in one frame, but vary in different frames. We use Jakes' model to represent the variation of the small-scale fading component for each channel in frame $t$, which yields \cite{liang2017spectrum}
\begin{align}
  \tilde{h}_m{(t)} &= \rho\tilde{h}_m{(t-1)} + e_{m}{(t)}, \label{eq:channelRelation1}\\
  \tilde{f}_n{(t)} &= \rho\tilde{f}_n{(t-1)} + e_{n}{(t)},\label{eq:channelRelation2}\\
  \tilde{g}_{m,n}{(t)} &= \rho\tilde{g}_{m,n}{(t-1)} + e_{m,n}{(t)},\label{eq:channelRelation3}
\end{align}
for $m = 1, \cdots, M$ and $n = 1, \cdots, N$, where $\tilde{h}_m{(0)}\sim\mathcal {CN}(0,1)$, $\tilde{f}_n{(0)}\sim\mathcal {CN}(0,1)$, and $\tilde{g}_{m,n}{(0)}\sim\mathcal {CN}(0,1)$, and $e_{m}{(t)}$, $e_{n}{(t)}$, and $e_{m,n}{(t)}$ are the independent and identically distributed random variables for any frame $t$ with distribution $\mathcal {CN}(0,1-\rho^2)$, and $\mathcal{CN}(\mu, \sigma^2)$ denotes the complex Gaussian distribution with mean $\mu$ and variance $\sigma^2$. The variable $\rho$ represents the correlation of channels between different frames.

\subsection{Signal Model}

As shown in Fig. \ref{fig:system model}, the BS transmits message $x_m$ with unit power to User $m$ in one time slot during one frame, while the IoT Device $n$ backscatters the received BS signals with its own message $c_n$ to one associated cellular user. Suppose that the symbol period for each IoT device covers $K$ BS symbol periods \cite{liu2013ambient}.
The received signals at User $m$ can be written as
\begin{equation}\label{eq:receivedSignal}
  y_m = \sqrt{p}h_mx_m+\sum_{n=1}^N a_{m,n}\sqrt{p}\alpha_n f_ng_{m,n}x_mc_n + u_m,
\end{equation}
where $p$ is the transmitted power at the BS, $\alpha_n$ denotes the reflection coefficient of IoT Device $n$, $u_m$ is the complex Gaussian noise at User $m$ with $u_m\sim\mathcal {CN}(0,\sigma^2)$, and $a_{m,n}\in\{0,1\}$ is the user association indicator. If $a_{m,n} = 1$, IoT Device $n$ is associated with User $m$, i.e., IoT Device $n$ transmits information when the BS serves User $m$; otherwise $a_{m,n} = 0$.

\subsection{SINR model}

The cellular user adopts SIC strategy to decode the messages for its own and the associated IoT devices. Due to the double fading, the backscatter link is weaker than the direct link. Thus, the cellular user needs to decode its own message first.
After that, the cellular user decodes the messages of the associated IoT devices. When there are multiple IoT devices are associated with the same cellular user, the cellular user first decodes the message of the strongest IoT device by treating other IoT devices' signals as interference. According to this strategy, we first define ${h}_{m,n} \triangleq |\alpha_n|^2|f_n|^2|g_{m,n}|^2$, and use set $\tilde{I}_n = \{l|h_{m,l}< h_{m,n}, l = 1,\cdots,N\}$ to indicate the identify numbers of the IoT devices that may interfere with IoT Device $n$. Then, the SINR of IoT Device $n$ at User $m$ is given by \cite{liang1999downlink}
\begin{eqnarray}
 \label{eq:SINRdmn}
\gamma_{m,n}=\frac{a_{m,n}Kp |\alpha_n|^2|f_n|^2|g_{m,n}|^2}{ \sum_{l\in \tilde{I}_n} a_{m,l} Kp{h}_{m,l} +\sigma^2}.
\end{eqnarray}

\section{Optimal User Association Policy}

In this section, we first formulate the user association problem for the SRN, which associates each IoT device with a suitable cellular user to maximize the sum rate of the IoT devices. Then we present the optimal policy for this formulated user association problem.

\subsection{Problem Formulation}
In SRN, the IoT transmission relies on the primary cellular transmission. Hence, different association scheme yields different IoT transmission rate, due to the different channel gains. Specifically, based on \eqref{eq:SINRdmn}, since each channel gain may vary in different frames, the IoT devices may need to be associated with different cellular users in different frames to achieve higher SINR, thereby higher transmission rate. Meanwhile, if there are multiple IoT devices associated with the same cellular user, there will exist interference which affects the IoT transmission rate.
Thus, it is significantly important to design a suitable user association policy. In what follows, we will formulate the user association problem mathematically.

The achievable rate, $R_{m,n}$, for IoT Device $n$ backscattering the signals to User $m$ is given by
\begin{eqnarray}
 \label{eq:ratemn}
R_{m,n}=\frac{1}{K}a_{m,n}\log_2(1+\gamma_{m,n}).
\end{eqnarray}
The sum rate for all IoT devices in the SRN can be written as $\sum_{m = 1}^{M}\sum_{n=1}^{N}R_{m,n}$.
Thus, the user association problem is expressed as
\begin{align}
 \label{eq:ProblemFormulation}
\mathbf{P1}:  \;\; \max\limits_{\mathbb{A}} &\;\;\sum_{m = 1}^{M}\sum_{n=1}^{N}R_{m,n}\\
s.t.&\;\;\sum_{m=1}^{M}a_{m,n} = 1, \label{eq:constraint}
\end{align}
where $\mathbb{A}$ represents the association index set composed by $a_{m,n}, m = 1,\cdots,M, n = 1,\cdots, N$, and \eqref{eq:constraint} means each IoT device only selects one time slot for information transmission in one frame, which is consistent with Fig.\ref{fig:frame}(b).

Note that the user association policy can be performed either at the IoT devices or at the BS. In this paper, we consider the user association policy is performed at the BS since the BS has stronger computing capacity than the IoT devices.

\subsection{The Optimal Policy} \label{sec:policy}

To obtain the optimal index set $\mathbb{A}^*$, it is clear that three steps are required: 1) list all possibility index set $\mathbb{A}$ satisfying \eqref{eq:constraint}; 2) calculate the sum rate of all IoT devices for each possible index set $\mathbb{A}$; 3) select the optimal index set $\mathbb{A}^*$ for maximizing $\sum_{m = 1}^{M}\sum_{n=1}^{N}R_{m,n}$.

It is noted that to solve the optimization problem $\mathbf{P1}$, the complete real-time channel information is required to calculate the real-time SINR in \eqref{eq:SINRdmn}. However, according to the frame structure in Fig. \ref{fig:frame}, each cellular user only receives the signals at its corresponding time slot, while each IoT device transmits its message only in one chosen time slot. Thus, the BS can only get the channel information from the IoT device to its associated cellular user. In other words, it is difficult for the BS to obtain the full real-time channel information from the IoT devices to all cellular users.
Therefore, it is impractical for the BS to calculate the optimal index set $\mathbb{A}^*$ and derive the optimal policy.


\section{Deep Reinforcement Learning Algorithms}

 In this section, we provide two DRL algorithms to solve the user association problem in SRN without requiring the full real-time channel information. One DRL algorithm, referred to as centralized DRL, uses the globally available information as the current states and obtains the user association decisions for all IoT devices at one time. The other DRL algorithm, referred to as distributed DRL, uses the locally available information as the current local states and obtains the user association decision only for one IoT device at one time. In the following, we will elaborate the basic principle, introduce the overview of DRL, and present the two DRL algorithms in detail.




\subsection{Basic Principle}\label{sec:basicPrinciple}

 As described in Section \ref{sec:policy}, the optimal policy of the user association problem requires the full real-time SINR in \eqref{eq:SINRdmn}, which means that the full real-time channel information is needed. However, it is impractical for the BS to obtain the full real-time channel estimation since in one frame, the BS can only obtain the channel information between the IoT device and its associated cellular user instead of all channel information between all IoT devices and all cellular users.

 In fact, the channels between different frames are correlated due to the following two reasons: 1) for the channels in different frames, the large-scale fading component remains constant if the location is unchanged; 2) the small-scale fading component follows the first-order complex Gauss-Markov process based on \eqref{eq:channelRelation1}, \eqref{eq:channelRelation2}, and \eqref{eq:channelRelation3}. Thus, if the BS can learn the correlation between the channels in different frames by exploring and exploiting the historical channel information, it is possible for BS to infer the current channel information and associate each IoT device with an appropriate cellular user to maximize the IoT sum transmission rate in each frame.

 DRL can effectively learn a hidden correlation by trial-and-error and design its optimal policy from the interaction with the environment \cite{mnih2015human}. Therefore, we can use DRL to learn the channel correction and design a proper user association policy to maximize the sum transmission rate.

\subsection{Overview of DRL}

 In this section, we will present the overview of DRL technology. For that, we first elaborate the RL framework.

\subsubsection{RL Framework}

\begin{figure}
\centering
\includegraphics[width=.88\columnwidth] {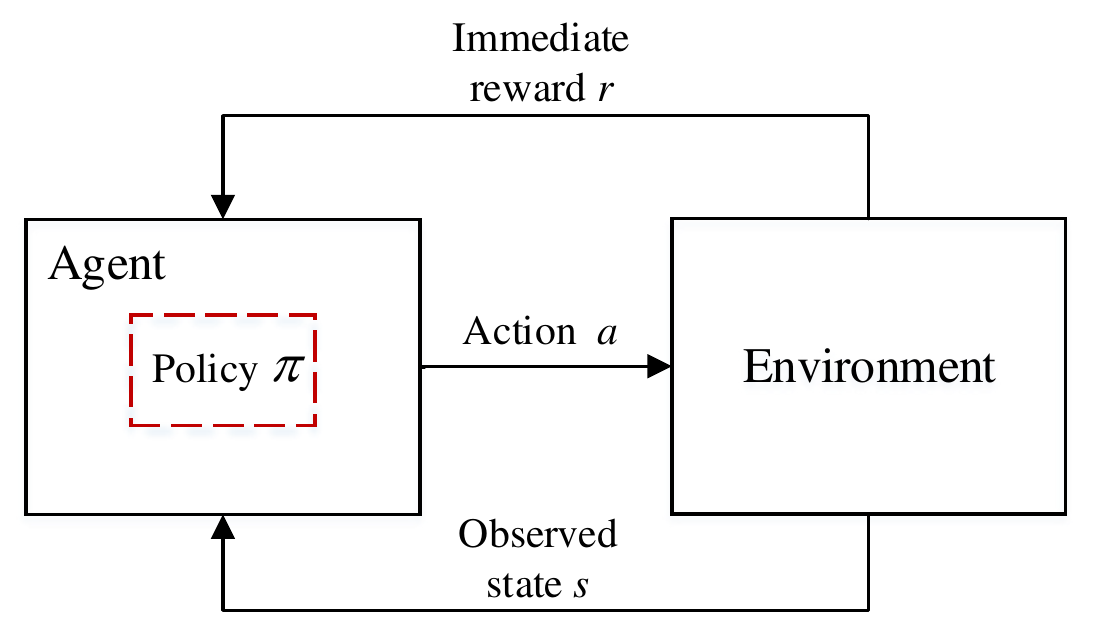}
\captionsetup{font={scriptsize}}
\caption{RL framework.}
\label{fig:RLframe}
\vspace{-0.5em}
\end{figure}

In reinforcement learning process, the agent learns its own best policy through interacting with its environment over time \cite{kaelbling1996reinforcement,sutton2018reinforcement}. Here, we first define the main elements of RL. Denote by $\mathcal{S}$ the set of all possible environment states $s$, by $\mathcal{A}$ the set of all possible actions $a$, by $r(s,a)$ the immediate reward when adopting action $a\in\mathcal{A}$ under the environment state $s \in \mathcal{S}$, and by $\pi$ the policy that the agent uses to map the current environment state to the pending action.

As shown in Fig. \ref{fig:RLframe}, the agent first observes the current state $s\in \mathcal{S}$, and then takes action $a \in \mathcal{A} $ by the current policy $\pi$. After taking an action, the environment state changes from $s$ to $s'\in \mathcal{S}$, and the agent gets an immediate reward $r(s,a)$. According to the observed information, $s'$ and $r(s,a)$, the agent repeatedly adjusts its policy to approach to the optimal policy.

The agent adjusts its policy to maximize the long-term reward. Notice that the maximization of the long-term reward is not equivalent to that of the immediate reward since for one state-action pair $(s,a)$ with a high immediate reward, its next state-action pair $(s',a')$ may suffer from a low immediate reward. Thus, the long-term reward includes not only the immediate reward but also the future reward, which can be expressed as
\begin{eqnarray}
 \label{eq:long-termReward}
\mathbf{\emph{Q}}(s,a)=r(s,a)+\gamma\sum_{s'\in \mathcal{S}}\sum_{a'\in\mathcal{A}}P_{s,s'}(a)\mathbf{\emph{Q}}(s',a'),
\end{eqnarray}
where $\gamma\in [0,1]$ is a discount factor indicating the impact of the future reward and $P_{s,s'}(a)$ denotes the transition probability from the state $s$ to the state $s'$ when taking action $a$. In the RL process, the agent aims to take an optimal action and find the optimal policy $\pi^*(s)$ under the current state $s$ by maximizing the long-term reward. Thus, based on \eqref{eq:long-termReward}, the optimal long-term reward $\mathbf{\emph{Q}}^*(s,a)$ can be written as
\begin{eqnarray}
 \label{eq:optimalLong-termReward}
\mathbf{\emph{Q}}^*(s,a)=r(s,a)+\gamma\sum_{s'\in \mathcal{S}}P_{s,s'}(a)\max_{a'\in\mathcal{A} }\mathbf{\emph{Q}}^*(s',a').
\end{eqnarray}
And the optimal policy $\pi^*(s)$ is
\begin{eqnarray}
 \label{eq:optimalPolicy}
\pi^*(s)=\arg \max_{a\in\mathcal{A} } \mathbf{\emph{Q}}^*(s,a).
\end{eqnarray}

Actually, the optimal policy in \eqref{eq:optimalPolicy} and the optimal long-term reward in \eqref{eq:optimalLong-termReward} are very difficult to be obtained directly since the transition probability $P_{s,s'}(a)$ is typically unknown for the agent especially with the complicated environment. The Q-learning algorithm is a well-known model-free RL algorithm to obtain the optimal policy, which does not require the transition probability $P_{s,s'}(a)$. Specifically, the Q-learning algorithm constructs a lookup $|\mathcal{S}|\times|\mathcal{A}|$ Q-table, in which $\mathbf{\emph{Q}}(s,a)$ as element indicates the long-term rewards of all possible state-action pairs. In addition, the agent takes actions through the $\epsilon$-greedy policy for each time step and obtains the corresponding experience $(s,a,r,s')$. After each experience $(s,a,r,s')$, the Q-learning algorithm updates the corresponding element in Q-table according to
\begin{equation}
 \label{eq:Qtable}
Q(s,a)\leftarrow (1-\alpha)Q(s,a) +\alpha\left[r(s,a)+\gamma\max_{a'\in\mathcal{A} }Q(s',a')\right],
\end{equation}
where $\alpha$ is the learning rate. Note that the Q-table is initialized randomly.

The $\epsilon$-greedy policy implies that the agent takes a random action from the action space $\mathcal{A}$ with probability $\epsilon$, whereas executes the action $a^*$ that makes the maximum value in the Q-table given a current state $s$, i.e., $a^* = \arg\max_{a\in\mathcal{A} }Q(s,a)$, with probability $1-\epsilon$ \cite{mnih2015human}. The $\epsilon$-greedy policy can avoid falling into the local optimum. The main reason is that the random action with a probability of $\epsilon$ can explore more possible action and experience the best action to update the Q-table.

In fact, when the state space and the action space are small, the Q-learning algorithm can rapidly experience all possible state-action pairs to update the Q-table, thereby high performance. However, in practice, the size of the state and action spaces are typically large, especially with complicated environment. In this case, the performance of the Q-learning algorithm is degraded since it is difficult to experience all possible actions especially the best action and it is unacceptable to storage the large Q-table. To overcome the shortcoming of the Q-learning algorithm, DRL is introduced to find the optimal policy under the large state-action spaces. In the following, we will provide the DRL framework.

\subsubsection{DRL Framework}

\begin{figure}
\centering
\includegraphics[width=.88\columnwidth] {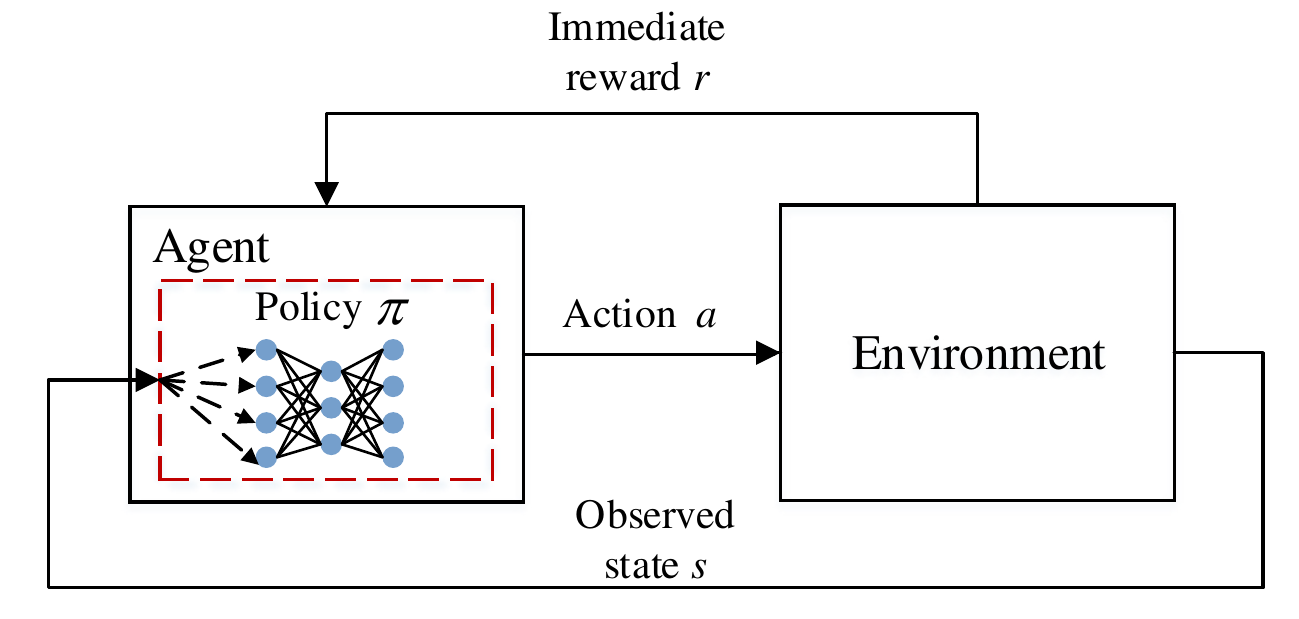}
\captionsetup{font={scriptsize}}
\caption{DRL framework.}
\label{fig:DRLframe}
\vspace{-0.5em}
\end{figure}

In DRL, a deep neural network, referred to as \emph{deep Q-network} (DQN), instead of the Q-table is implemented to estimate the long-term reward $\mathbf{\emph{Q}}(s,a)$, as shown in Fig. \ref{fig:DRLframe}. The DQN can be expressed as $\mathbf{\emph{Q}}(s,a;\bm \theta)$, where $\bm \theta$ is the weights of the DQN. The input of the DQN is one of the environment states, i.e., $s\in \mathcal{S}$, and the output is the long-term reward $\mathbf{\emph{Q}}(s,a;\bm \theta)$ of each possible action $a$ in $\mathcal{A}$ for a given environment state $s$. In fact, for DRL, to achieve an approximate value $Q^*(s,a)$, the agent needs to update the DQN weights $\bm \theta$, which is equivalent to the update of Q-table in RL. Similarly, the DRL uses each experience $(s,a,r,s')$ obtained by the $\epsilon$-greedy policy to train the DQN. The process of training DQN aims to minimize the loss function $\mathbb{L(\bm \theta)}$, which can be expressed as
\begin{equation}
 \label{eq:LossFunction}
\mathbb{L(\bm \theta)} = \mathbb{E} \left[|y_{tar}-\mathbf{\emph{Q}}(s,a;\bm \theta)|^2\right],
\end{equation}
where $y_{tar}$ is the target value, which is given by
\begin{equation}
 \label{eq:TargetValue}
y_{tar} = r(s,a) + \gamma\max_{a'\in\mathcal{A} }  \mathbf{\emph{Q}}(s',a';\bm \theta^-),
\end{equation}
 where $\bm \theta^-$ is the old weights of the DQN, which is updated once per $T_u$ steps. We call $\mathbf{\emph{Q}}(s,a;\bm \theta^-)$ the target Q-network, which updates its weights $\bm \theta^-$ frequently but slowly. The target Q-network can stabilize the learning algorithm by removing the correlations among the targets and the estimated Q-values.

 Note that in DQN, experience replay mechanism is also used to overcome the instability of the learning algorithm \cite{mnih2015human}. During the learning process, the agent not only uses the current experience $(s,a,r,s')$, but also uses the old experiences. In particular, the neural network is trained by randomly sampling a minibatches of $Z$ experiences from the replay memory $\mathbb{D}$. The replay memory $\mathbb{D}$ is used to store the experiences $(s,a,r,s')$ with a first-in-first-out principle. Once getting a new experience, the agent puts it into the replay memory $\mathbb{D}$. The size of this replay memory $\mathbb{D}$ is $N_E$. By using the experience replay mechanism, the experiences used for learning are more like independent and identically distributed, thereby reducing the correlations among the observations. Therefore, the experience replay mechanism increases the stability of the learning process.

\subsection{Centralized DRL-based User Association Algorithm}\label{sec:centrolized}

 In this subsection, we present the centralized DRL-based user association algorithm, in which the BS serves as the agent. In this algorithm, the BS makes the user association decisions for all IoT devices at one time for a given environment state. To begin with, we introduce the action space, the state space, and the immediate reward function for this algorithm.

\subsubsection{Actions}

Since the centralized DRL algorithm aims to associate each IoT device with a proper cellular user to maximize the sum rate, the action space needs to include all possible and available association schemes. Thus, the action space is given by
\begin{equation}
 \label{eq:C_action}
\mathcal{A}_c = \{\{b_1,\cdots, b_N\}_1,\cdots, \{b_1,\cdots, b_N\}_{M^N}\},
\end{equation}
where $b_n\in\{1, \cdots, M\}$ denotes the index of the cellular user associated with the IoT Device $n$. The number of possible actions is $M^N$, i.e., the size of this action space is $M^N$. We take an example to understand this action space. Assuming that there are $M = 2$ cellular users and $N = 2$ IoT devices, the action space $\mathcal{A}_c$ is $\mathcal{A}_c = \{\{1,1\},\{1,2\},\{2,1\},\{2,2\}\}$, which means there are $2^2 = 4$ possible actions.

\subsubsection{States}\label{sec:C_state}

Since the DRL agent trains the DQN based on each experience $(s,a,r,s')$, it is important for the DRL agent to collect a proper and available state to provide useful knowledge for decision making.
In Section \ref{sec:policy}, we have stated that the full real-time channel information is difficult to be obtained. However, the channels between different frames are correlated, which has been discussed in Section \ref{sec:basicPrinciple}. As such, we can use the historical channel information as the state to optimize the policy.

Here, we denote by $\mathcal{H}_{L} = \{h_{m,n}\}$ the historical channel information of all backscatter links. After each interaction with environment, $\mathcal{H}_{L}$ will be update. In particularly, at the end of frame $t$, User $m$ transmits the backscatter channels information $h_{m,n}(t)$ from IoT Device $n$ associated with it to the BS. Then the BS updates $\mathcal{H}_{L}(t)$ with the information $h_{m,n}(t)$ and considers the updated $\mathcal{H}_{L}(t)$ as the state for the frame $(t+1)$. To summarize, the state in frame $t$ is given by
\begin{equation}
 \label{eq:C_state}
s_c(t) = \mathcal{H}_{L}(t-1).
\end{equation}
Note that before feeding $s_c(t)$ into DQN, we first normalize it to guarantee the performance of the centralized DRL algorithm.

\subsubsection{Reward Function}

The goal of this centralized DRL algorithm is to maximize the sum rate of all IoT devices. Thus the immediate reward function $r_c(t)$ in frame $t$ shall be the sum rate of all IoT devices. i.e.,
\begin{equation}
 \label{eq:C_reward}
r_c(t) = \sum_{m=1}^{M}\sum_{n=1}^{N}R_{m,n}(t).
\end{equation}
Note that after taking action $a$ by the observed state $s$, the BS will obtain the immediate reward by the feedback from the cellular users.

\begin{figure}
\centering
\includegraphics[width=.88\columnwidth] {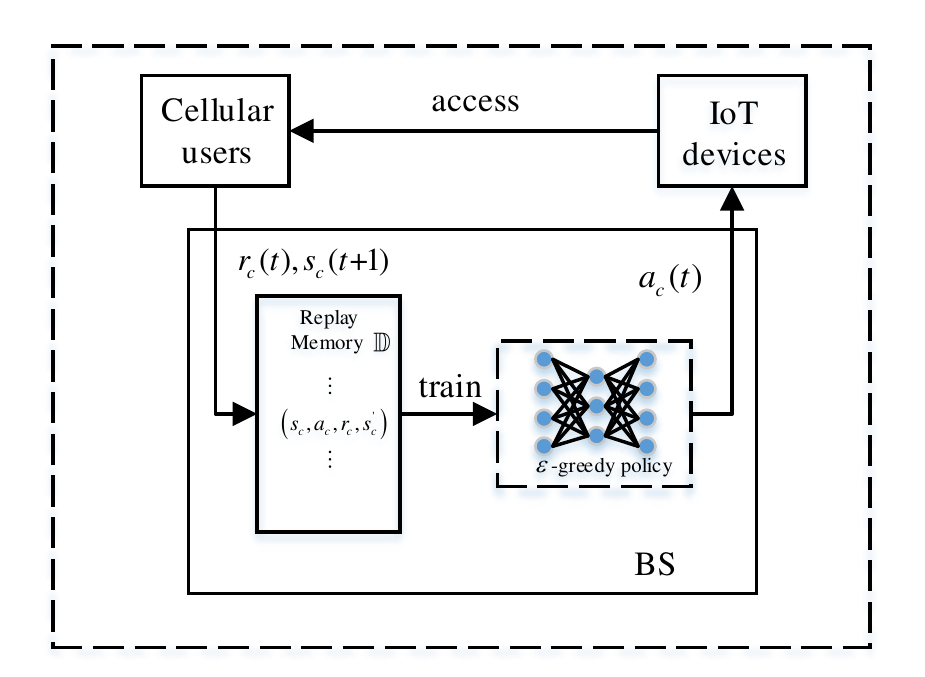}
\captionsetup{font={scriptsize}}
\caption{The structure of the proposed centralized DRL-based user association algorithm.}
\label{fig:C_structureDRL}
\vspace{-0.5em}
\end{figure}

Fig. \ref{fig:C_structureDRL} shows the structure of the proposed centralized DRL algorithm. In this algorithm, the agent delivers the decision $a_c(t)$ made according to $\epsilon$-greedy policy to the IoT devices. The IoT devices access the associated cellular user based on the decision from the BS. And the cellular users decode the signals of the associated IoT devices and feedback all useful and available information to BS for the calculation and the update of $r_c(t)$ and $s_c(t+1)$. Then the BS storages the experience $(s_c(t),a_c(t),r_c(t),s_c(t+1))$ into the replay memory $\mathbb{D}$, and randomly samples a minibatch of experiences in $\mathbb{D}$ to train the DQN. The DQN is used to make decision for the next frame according to $\epsilon$-greedy policy.
In addition, the pseudocode of the proposed centralized DRL-based user association algorithm is shown in Algorithm \ref{alg:centralized}.

\begin{algorithm}[h]
  \caption{Centralized DRL-based User Association Algorithm}
  \begin{algorithmic}[1]\label{alg:centralized}
   \STATE  Initialize the weights $\bm \theta_c$ of the DQN randomly;\\
   \STATE  Initialize the weights $\bm \theta^-_c$ of the target Q-network with $\bm \theta^-_c = \bm \theta_c$;\\
   \STATE  Initialize the size of minibatch $Z$;\\
   \STATE  Initialize the target Q-network replacement frequency $T_u$;\\
   \STATE  The agent takes actions randomly and storages the corresponding experience $(s_c,a_c,r_c,s_c')$ into the replay memory $\mathbb{D}$ until there are $Z$ experiences.
   \STATE  {\bf{Repeat:}}\\
      \STATE  The agent selects an action $a_c(t)$ through $\epsilon$-greedy policy in frame $t, (t>Z)$;\\
      \STATE  The agent calculates the immediate reward $r_c(t)$ after taking action $a_c(t)$ in frame $t$;\\
      \STATE  The agent observes a new state $s_c(t+1)$ in frame $(t+1)$;\\
      \STATE  The agent stores the new experience $(s_c(t),a_c(t), r_c(t), s_c(t+1))$ into the replay memory $\mathbb{D}$;\\
      \STATE  The agent randomly samples a minibatch of $Z$ experiences $(s_c,a_c,r_c,s_c')$ from the replay memory $\mathbb{D}$ to train the DQN;\\
      \STATE  The agent updates the DQN weights $\bm \theta_c$;\\
      \STATE  The agent updates the target Q-network weights $\bm \theta^-_c$ once per $T_u$ steps with $\bm \theta^-_c = \bm \theta_c$.\\
\end{algorithmic}
\end{algorithm}

For the centralized DRL algorithm, when $N$ is large, the state-action space becomes very large. In this case, it is difficult for this algorithm to train the DQN successfully. In addition, if $N$ increases, this algorithm can not work since the state-action space changes, resulting in the inability to use the designed DQN. In other words, the centralized DRL algorithm is not a scalable algorithm. To overcome the above challenges, we propose another algorithm called distributed DRL-based user association algorithm. In the following, we will present this algorithm.

\subsection{Distributed DRL-based User Association Algorithm}

In this subsection, we provide a distributed DRL-based user association algorithm, in which the BS serves as the agent and uses a centralized training and distributed execution framework \cite{calabrese2016learning}. In this algorithm, the BS allocates $N$ computing units to make decision for $N$ IoT devices individually. In other words, Unit $n$ inputs the state of IoT Device $n$ and outputs the action of IoT Device $n$.

Here, we first introduce the action space, the state space, and the immediate reward function.

\subsubsection{Actions}

In the distributed DRL algorithm, the computing unit makes decision only for one IoT device at one time with a given state of the corresponding IoT device. Thus, the action space is given by
\begin{equation}
 \label{eq:D_action}
\mathcal{A}_d = \{1,2,\cdots, M\}.
\end{equation}

\subsubsection{States}

Since the units make the user association decision individually, it is difficult to control the decision of other unit in the distributed DRL algorithm. We notice that the optimal user association policy for this algorithm is not only related to the channel information but also related to the interference information. This means that the state requires not only the historical channel information, but also the interference information. The interference information includes two components: interferer information and interfered information.
In what follows, for the distributed DRL algorithm, we describe the state $s_d^{n}(t)$ of IoT Device $n$ conditioned on associated with User $m$ at frame $t$, which is divided into three feature groups.
\begin{itemize}
  \item Local Information: According to \eqref{eq:SINRdmn}, the agent needs to feed the channel information into the DQN to provide useful knowledge for learning the optimal policy. Since it is difficult to obtain the channel information between IoT Device $n$ and all cellular users, the agent uses the historical information to explore and infer the current channel information, which is similar to Section \ref{sec:C_state}. Thus, the state is designed to include the historical channel information $\mathcal{H}_L^n(t-1)$, where $\mathcal{H}_L^n(t-1)$ is the historical channel information between IoT Device $n$ and all cellular users updated in frame $(t-1)$. Meanwhile, the state at frame $t$ includes the action taking by IoT Device $n$ at frame $(t-1)$ to suggest the effect of historical action. In addition, since the agent trains the DQN using all environment experiences, in order to identify all IoT devices, the state is designed to include the identity number, i.e., $n$.


  \item Interferer Information: The state is designed to include the interferer information to observe the interference from other IoT devices when decoding the message from IoT Device $n$. In particular, according to \eqref{eq:SINRdmn}, if IoT Device $n$ is associated with User $m$ in frame $(t-1)$, User $m$ will feedback the interferer information when decoding the IoT Device $n$ message, $I_n(t-1)$, to the BS, where $I_n(t-1) = \sum_{l\in \tilde{I}_n(t-1)}p a_{m,l}^{d}(t-1){{h}}_{m,l}(t-1)$ and $a_{m,l}^{d}(t-1)\in \{0,1\}$ indicates whether the IoT Device $l$ is associated with User $m$ in frame $(t-1)$.

  \item Interfered Information: Finally, the agent uses the feedback from User $m$ to sense the interference $O_n(t-1)$ from IoT Device $n$ to other IoT devices in frame $(t-1)$, where $O_n(t-1) = \sum_{l\in \tilde{O}_n(t-1)}p a_{m,l}^{d}(t-1){{h}}_{m,l}(t-1)$ and $\tilde{O}_n(t-1) =  \{l|h_{m,l}(t-1)> h_{m,n}(t-1), l = 1,\cdots,N\}$ is the identify number set of the IoT device that may be interfered by IoT Device $n$ at frame $(t-1)$. And the state is designed to include the interference information $O_n(t-1)$.

\end{itemize}

To summarize, the state $s_d^{n}(t)$ of IoT Device $n$ at frame $t$ is given by
\begin{equation}
 \label{eq:D_state}
s_d^{n}(t) = \{\mathcal{H}_L^n(t-1),a_d^n(t-1), n, I_n(t-1), O_n(t-1) \},
\end{equation}
where $a_d^n(t-1)$ is the action of IoT Device $n$ at frame $(t-1)$. Notice that $s_d^{n}(t)$ is normalized to guarantee the performance of the distributed DRL algorithm.

\subsubsection{Reward Function}

The immediate reward function should evaluate the effect of the action taken on the goal of maximizing the sum rate. Here, the immediate reward includes not only the current transmission rate of IoT Device $n$, but also the interference with other IoT devices. The main reason is that if the decision process for each IoT device aims to maximize its own transmission rate, it is difficult to converge to an optimal policy for maximizing the sum rate.

To quantify the effect of interference, similar to \cite{nasir2018deep}, the agent first calculates the transmission rate, $R_{m,l}^{-n}(t)$, without the interference from IoT Device $n$ for IoT Device $l\in \tilde{O}_n(t)$, which can be expressed as
\begin{align}
 \label{eq:D_withoutInterference}
&R_{m,l}^{-n}(t)= \frac{1}{K}a_{m,l}^{d}(t)\times \nonumber \\
&~~~~~~\log_2\left(1+\frac{a_{m,l}^{d}(t)Kp |\alpha_l|^2|f_l|^2|g_{m,l}|^2}{ \sum_{i\in\tilde{I}_l(t) ,i\neq n} a_{m,i}^{d}(t) Kp{{h}}_{m,i} +\sigma^2}\right),
\end{align}
Then, the agent computes the effect of IoT Device $n$ on the IoT devices in $\tilde{O}_n(t)$ by
\begin{equation}
 \label{eq:D_interferenceEffect}
\beta_{l}^{-n}(t) = R_{m,l}^{-n}(t)- R_{m,l}(t).
\end{equation}
Thus the immediate reward function can be written as
\begin{equation}
 \label{eq:D_reward}
r_d^n(t) = R_{m,n}(t)- \sum_{l\in \tilde{O}_n(t)}\beta_{l}^{-n}(t).
\end{equation}
The reward in \eqref{eq:D_reward} consists of two components: its contribution to the sum rate and the penalty about interference to other IoT device. This reward function ensures that the agent considers not only the maximization of each IoT device rate, but also the effect on other IoT devices, thereby guaranteeing the optimal policy rapidly.

\begin{figure}
\centering
\includegraphics[width=.88\columnwidth] {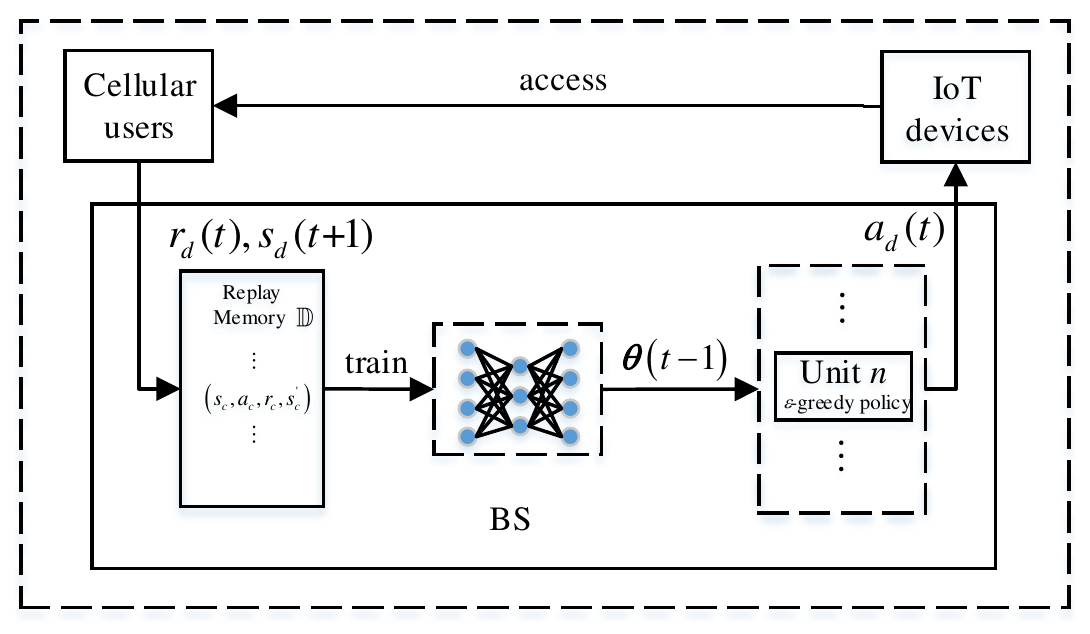}
\captionsetup{font={scriptsize}}
\caption{The structure of the proposed distributed DRL-based user association algorithm.}
\label{fig:D_structureDRL}
\vspace{-0.5em}
\end{figure}

The structure of the proposed distributed DRL algorithm is shown in Fig. \ref{fig:D_structureDRL}. The information delivery between cellular users, the IoT devices, and the BS is the same with the centralized DRL algorithm, which is discussed in Section \ref{sec:centrolized}. The difference between these two algorithms is that the BS needs to allocate $N$ computing units to make decisions for $N$ IoT devices individually in the distributed DRL algorithm. In addition, in the distributed DRL algorithm, after training the DQN, the BS delivers the updated DQN weights $\bm \theta_d$ to each computing unit. Then, the $N$ computing units make decisions, individually, for the $N$ IoT devices according to the $\epsilon$-greedy policy. In addition, the pseudocode of the proposed distributed DRL-based user association algorithm is shown in Algorithm \ref{alg:distributed}.

Note that if the number of IoT devices $N$ changes, the BS just changes the number of computing units to execute the distributed DRL algorithm without redesigning the DQN. In other words, this distributed DRL algorithm has the advantage of scalability. In addition, here, we consider the units at the BS make decisions for the IoT devices due to the limited computing capability of the IoT devices. If the IoT devices have enough computing capacity, it is reasonable that each IoT device as the agent makes its own decisions based on the proposed distributed DRL algorithm.

\begin{algorithm}[h]
  \caption{Distributed DRL-based User Association Algorithm}
  \begin{algorithmic}[1]\label{alg:distributed}
   \STATE  Initialize the weights $\bm \theta_d$ of the DQN randomly;\\
   \STATE  Initialize the weights $\bm \theta^-_d$ of the target Q-network with $\bm \theta^-_d = \bm \theta_d$;\\
   \STATE  Initialize the size of minibatch $Z$;\\
   \STATE  Initialize the target Q-network replacement frequency $T_u$;\\
   \STATE  The agent takes actions randomly and storages the corresponding experience $(s^n_d,a^n_d,r^n_d,(s^n_d)')$ of each IoT device into the replay memory $\mathbb{D}$ until there are $Z$ experiences.
   \STATE  {\bf{Repeat:}}\\
      \STATE  Unit $n$ selects an action $a_d^n(t)$ through $\epsilon$-greedy policy in frame $t, (t>Z)$ for IoT Device $n$, $n = 1, \cdots, N$;\\
      \STATE  Unit $n$ calculates the immediate reward $r_d^n(t)$ after taking action $a_d^n(t)$ in frame $t$ for IoT Device $n$, $n = 1, \cdots, N$;\\
      \STATE  The agent observes a new state $s_d^n(t+1)$ in frame $(t+1)$ of IoT Device $n$, $n = 1, \cdots, N$;\\
      \STATE  The agent stores all new experiences $(s_d^n(t),a_d^n(t), r_d^n(t), s_d^n(t+1))$, $n = 1,\cdots, N$ into the replay memory $\mathbb{D}$;\\
      \STATE  The agent randomly samples a minibatch of $Z$ experiences from the replay memory $\mathbb{D}$ to train the DQN;\\
      \STATE  The agent updates the DQN weights $\bm \theta_d$;\\
      \STATE  The agent updates the target Q-network weights $\bm \theta^-_d$ once per $T_u$ steps with $\bm \theta^-_d = \bm \theta_d$;\\
      \STATE  The agent delivers the updated DQN weights $\bm \theta_d$ to $N$ computing units.\\
\end{algorithmic}
\end{algorithm}

\section{Performance Evaluation}
\label{sec:simulations}

In this section, simulation results are presented to evaluate the performance of the two proposed DRL-based user association algorithms.
For comparison, we consider two benchmark algorithms: random policy and optimal policy. In the random policy, each IoT device is associated with a cellular user randomly. For the optimal policy, we assume that the BS knows full perfect real-time channel information and obtains the optimal policy by the method proposed in Section \ref{sec:policy}. Since it is impractical for the BS to perfectly know the full real-time channel information, the performance of the optimal policy is just the theoretical upper bound. In the following, we will present the simulation setup and the performance of the two proposed DRL-based user association algorithms.

\subsection{Simulation Setup}

To begin with, we consider the locations of the BS, the cellular users, and the IoT devices, are in a $100$ meters by $100$ meters region. The BS is located at the center of this region. And the IoT devices and the cellular users are placed randomly based on a uniform distribution within a distance of $10\thicksim 100$ meters from the BS.

We set the transmit power of the BS to $p = 40$dBm and the background noise power to $\sigma^2 = -114$dBm. We consider a distance-dependent path loss model, which is $32.45+20\log_{10}(f)+20\log_{10}(d)-G_t-G_r$ (in dB), where $f$ is the carrier frequency in Mhz, $d$ is the distance in km, $G_t$ denotes the transmit antenna gain, and $G_r$ denotes the receive antenna gain. Here we set $f = 2.4$GHz, $G_t = G_r = 2.5$dB. We assume all IoT devices have the same reflection coefficient $\alpha_n = \alpha = 0.8$ for $n = 1,\cdots, N$. And the period ratio between the IoT device and the BS is set to $K = 50$.

\begin{table}\label{tab:central}
\center
\caption{ Parameter Design for the Centralized DQN (C-DQN) and the Distributed DQN (D-DQN).}
\begin{tabular}{c c}
\hline
Parameters & Value \\
\hline
C-DQN: number of hidden layers & $3$ \\
C-DQN: neuron network size  & $256\times128\times64$ \\
D-DQN: number of hidden layers & $3$ \\
D-DQN: neuron network size  & $128\times64\times32$ \\
Activation function & ReLU \\
Optimizer & Adam \\
Learning rate & $0.01$ \\
Mini-batch size ($Z$) & $64$ \\
Replay memory size ($N_E$) & $800$ \\
Target-DQN updating frequency ($T_u$) & $100$ \\
\hline
\vspace{-1.8em}
\end{tabular}
\end{table}

Next, we describe the design of the hyper-parameters for the two DRL algorithms. First, the two DRL algorithms are implemented using TensorFlow, and the parameters of the two DQNs corresponding to the two DRL algorithms are listed in Table I. Furthermore, we set the discount factor to $\gamma = 0.3$. In addition, the $\epsilon$-greedy policy is used to take actions. At first, we set $\epsilon(0) = 0.2$, which means a random action is chosen with a probability of $0.2$ to explore the experiences. Then, to move from a more explorative policy to a more exploitative policy, the probability $\epsilon$ follows $\epsilon(t+1) = \max\{\epsilon_{\min},(1-\lambda_\epsilon)\epsilon(t)\}$, where $\epsilon_{\min} = 0.005$ and $\lambda_\epsilon = 0.005$.

\subsection{Performance for the Proposed Algorithms}

\begin{figure}[t]
\centering
  \begin{subfigure}[b]{.99\linewidth}\label{fig:1a}
    \centering
    \includegraphics[width=8.5cm]{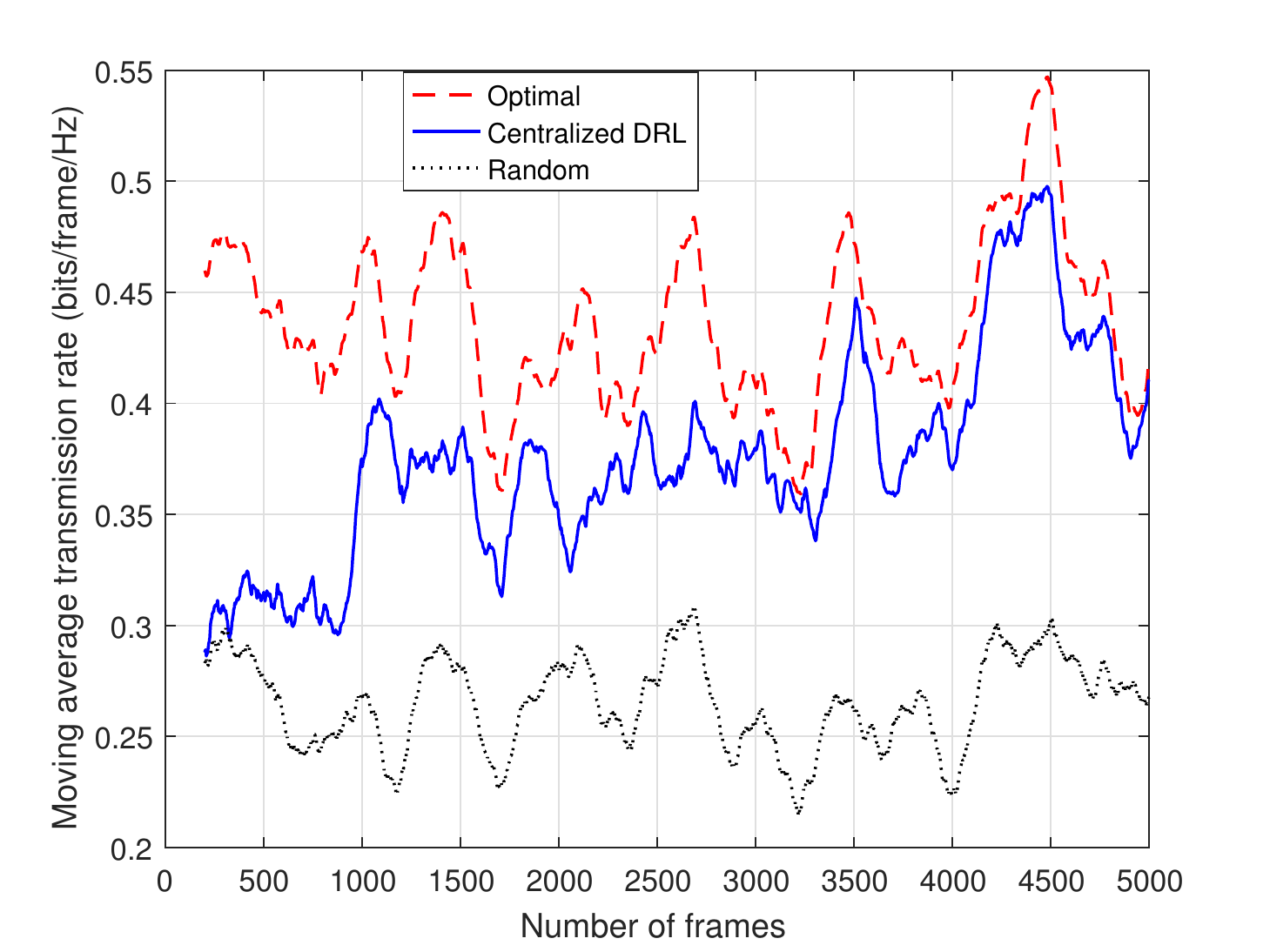}
    \centering
    \caption{}
  \end{subfigure}\\ %
  \begin{subfigure}[b]{.99\linewidth}\label{fig:1b}
    \centering
    \includegraphics[width=8.5cm]{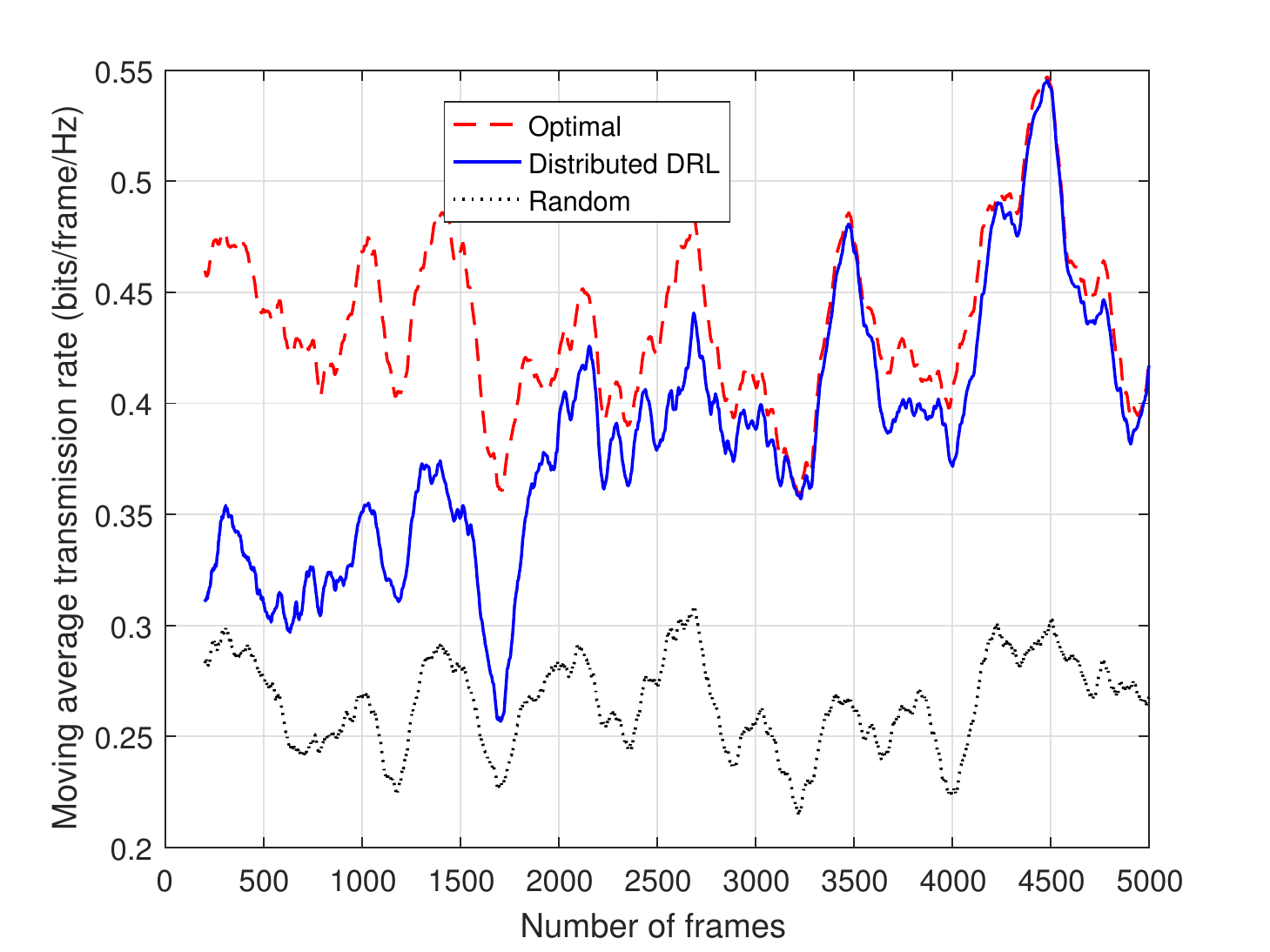}
    \centering
    \caption{}
  \end{subfigure}
  \captionsetup{font={scriptsize}}
  \caption{The average sum transmission rate comparison with $\rho = 0.99$. Each value is a moving average of the pervious $200$ frames.}\label{fig:DRL1}
\vspace{-2em}
\end{figure}

Fig. \ref{fig:DRL1} illustrates the average sum transmission rate of all IoT devices using different algorithms. In this figure, we consider a quasi-static channel scenario by setting $\rho = 0.99$, which means the channel changes slowly. Meanwhile, we consider the number of the cellular users is $M = 3$ and the number of the IoT devices is $N = 3$ in this figure. It can be seen that both the centralized DRL algorithm and the distributed DRL algorithm can almost achieve the optimal sum transmission rate gradually in a quasi-static scenario. This observation indicates that the two DRL algorithms can learn almost perfect knowledge and design almost optimal policy in a quasi-static scenario. Meanwhile the average sum rate of the proposed two DRL algorithms is around $0.45$ bits/frame/Hz, while the average sum rate of the random policy is around $0.25$ bits/frame/Hz. That indicates the average sum rate of the proposed two DRL algorithms is almost twice the average sum rate of random policy.

\begin{figure}
\centering
\includegraphics[width=.99\columnwidth] {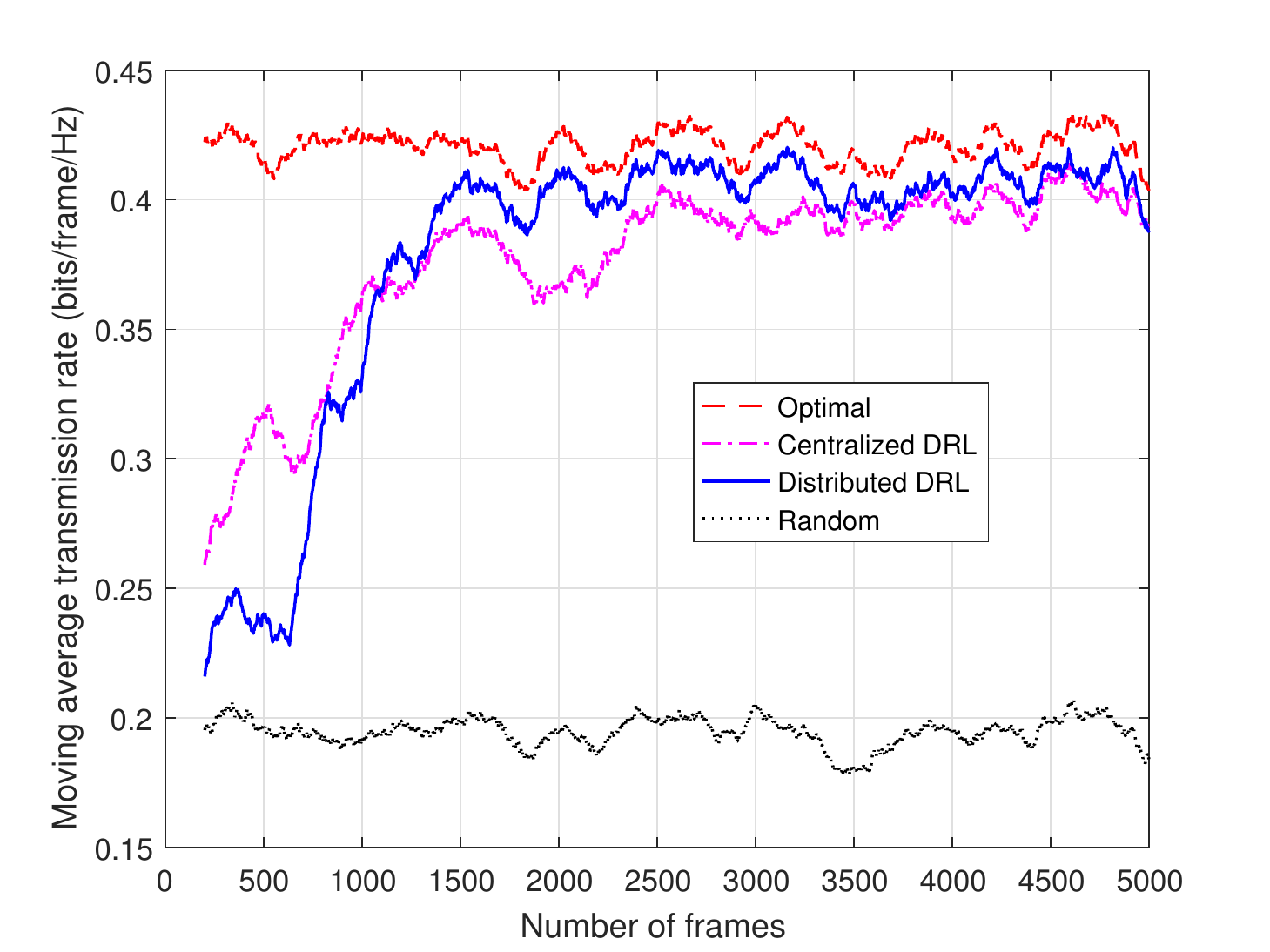}
\captionsetup{font={scriptsize}}
\caption{The average sum transmission rate comparison with $\rho = 0.5$. Each value is a moving average of the pervious $200$ frames.}
\label{fig:DRL5}
\vspace{-0.5em}
\end{figure}

Fig. \ref{fig:DRL5} presents the average sum transmission rate for different algorithms in a relative dynamic channel scenario with $\rho = 0.5$. In this figure, we set $M = N = 3$. From this figure, we can see that the two DRL algorithms can approach the performance of the optimal policy. Compared with the quasi-static scenario, the two DRL algorithms in this more dynamic scenario have a little gaps with the optimal policy. The main reason is that when the channel changes rapidly, it is more difficult to infer the next channel state. 


\begin{figure}
\centering
\includegraphics[width=.99\columnwidth] {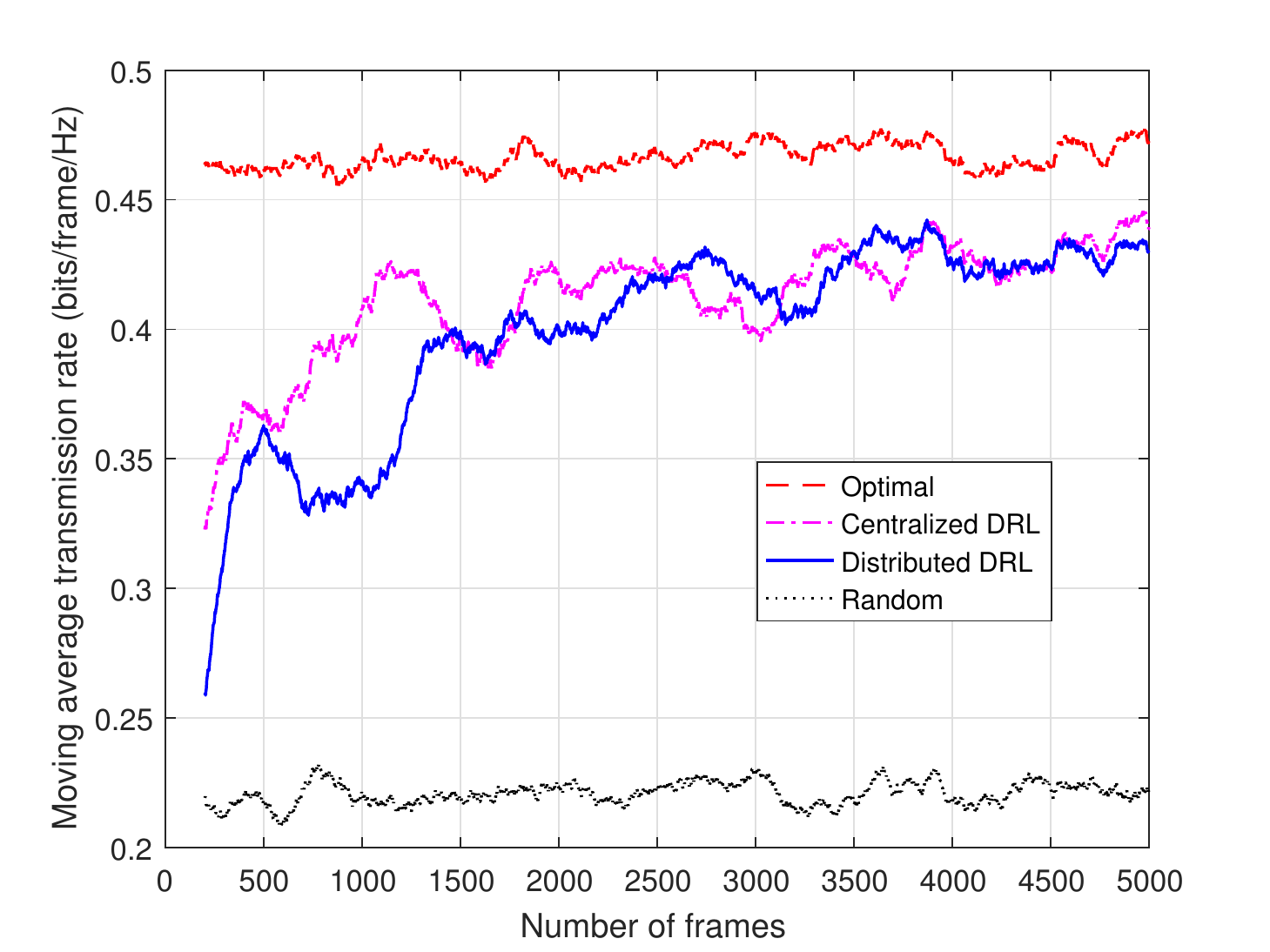}
\captionsetup{font={scriptsize}}
\caption{The average sum transmission rate comparison with $\rho = 0$. Each value is a moving average of the pervious $200$ frames.}
\label{fig:DRL0}
\vspace{-0.5em}
\end{figure}

Fig. \ref{fig:DRL0} shows the performance of the average sum transmission rate for different algorithms in a highly dynamic scenario with $\rho = 0$. In this scenario, the small-scale fading component changes rapidly without correlation between different frames. We set $M = N = 3$.
From this figure, it is seen that there exist gaps between the two proposed DRL algorithms and the optimal policy. The main reason is that when $\rho = 0$, the small-scale fading component is difficult to be learnt from the historical channel information since the channel changes without correlation between different frames. However, the proposed algorithms can approach the optimal policy. This is because the agent can learn the large-scale fading information. These observations indicate that the proposed two algorithms are effective even in a highly dynamic scenario.

\subsection{The Scalability of the Distributed DRL Algorithm}

\begin{figure}[t]
\centering
  \begin{subfigure}[b]{.9\linewidth}\label{fig:de}
    \centering
    \includegraphics[width=8.5cm]{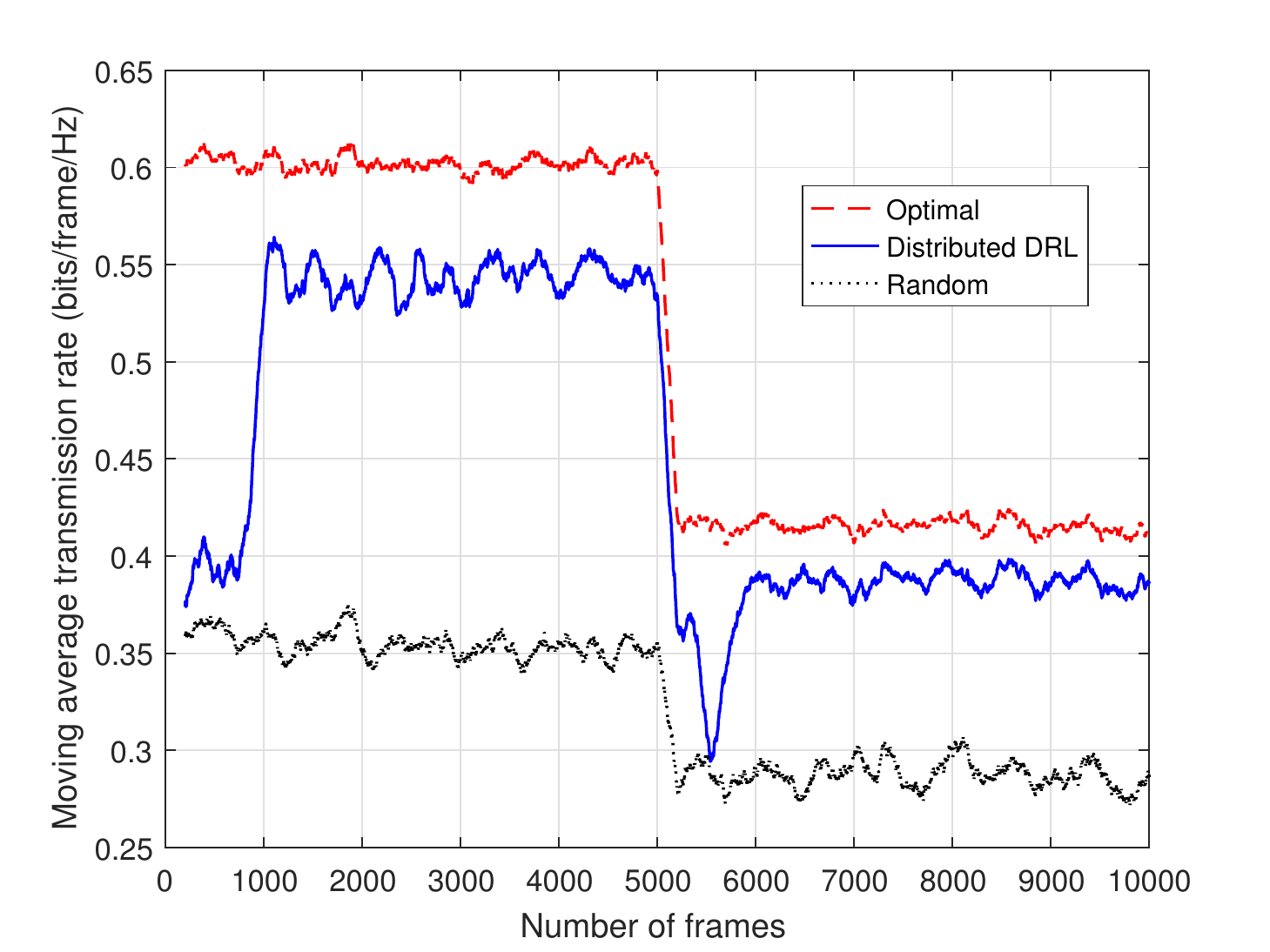}
    \centering
    \caption{The number of IoT devices changes from $N = 3$ to $N = 2$.}
  \end{subfigure}\\
  \begin{subfigure}[b]{.9\linewidth}\label{fig:in}
    \centering
    \includegraphics[width=8.5cm]{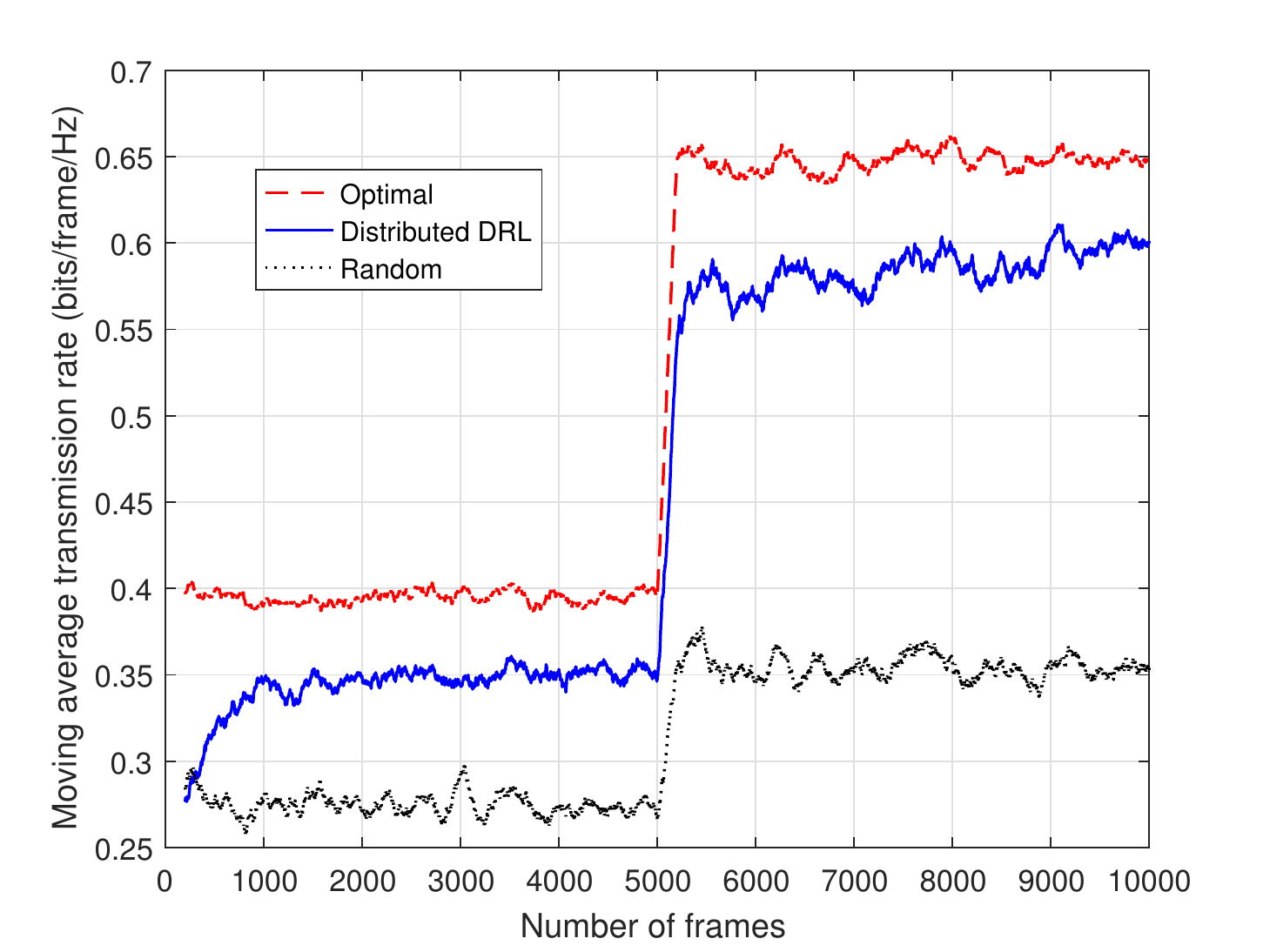}
    \centering
    \caption{The number of IoT devices changes from $N = 2$ to $N = 3$.}
  \end{subfigure}
  \captionsetup{font={scriptsize}}
  \caption{The average sum transmission rate comparison with $\rho = 0$. Each value is a moving average of the pervious $200$ frames.}\label{fig:DDRL5}
\vspace{-2em}
\end{figure}

In this subsection, we present the scalability of the proposed distributed DRL algorithm. When the number of IoT devices changes, the centralized DRL algorithm can not work effectively due to the change of the action space. Fig. \ref{fig:DDRL5} shows the performance of the average sum rate of different algorithms when the number of IoT devices changes with $\rho = 0$ and $M = 3$. It is seen that regardless of whether the number of IoT devices increases or decreases, the distributed DRL algorithm can approach the optimal policy, and always be better than the random policy. This figure validates the scalability of the proposed distributed DRL algorithm when the environment changes in a highly dynamic way.

\begin{figure}
\centering
\includegraphics[width=.99\columnwidth] {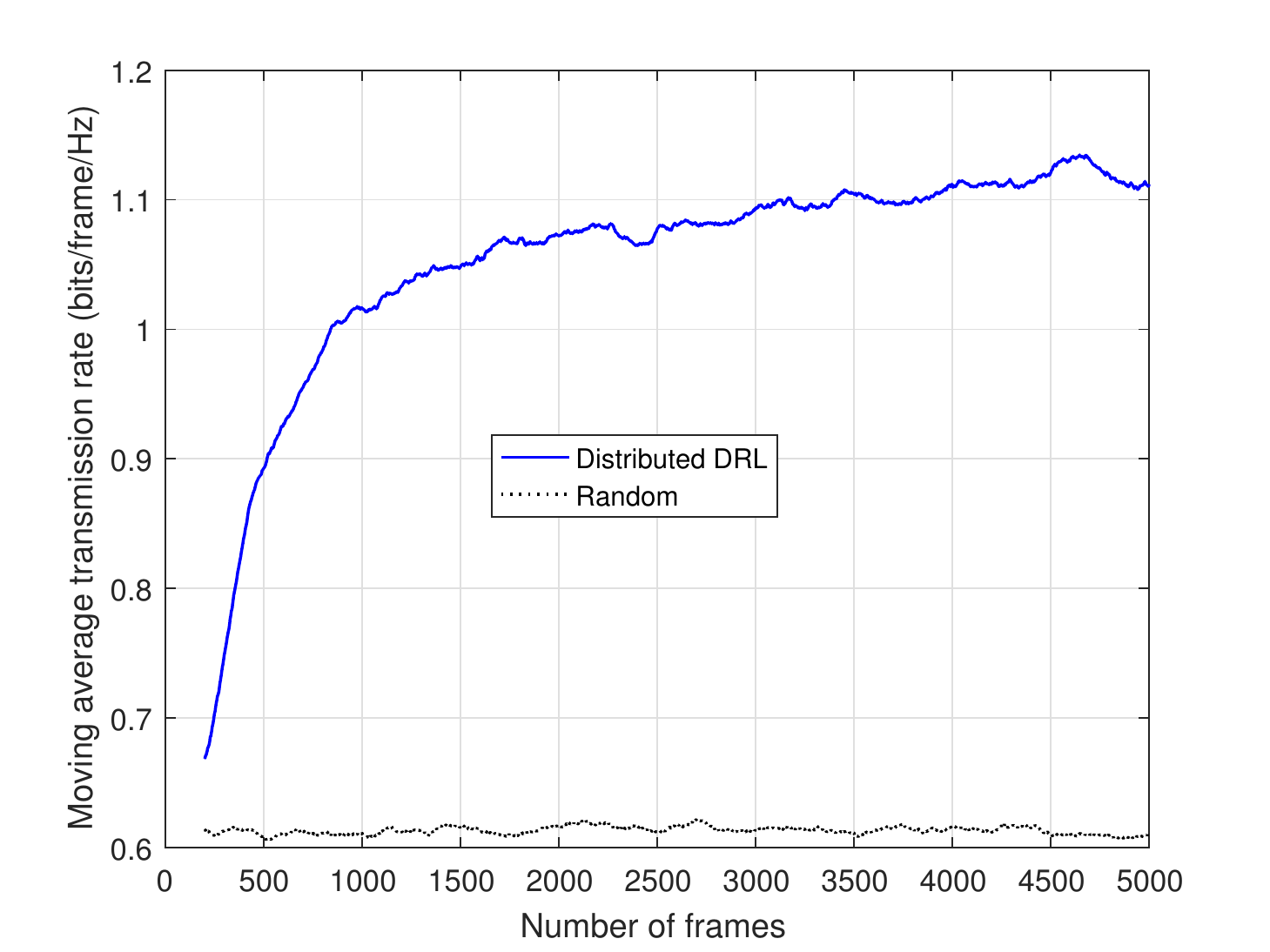}
\captionsetup{font={scriptsize}}
\caption{The average sum transmission rate comparison with $\rho = 0.5$. Each value is a moving average of the pervious $200$ frames.}
\label{fig:DDRLi}
\vspace{-0.5em}
\end{figure}

Fig. \ref{fig:DDRLi} depicts the performance of the average sum rate for different algorithms with $\rho = 0.5$, $M = 8$, and $N = 8$. In this case, the size of action space for the centralized DRL algorithm is $8^8 = 1.68\times10^7$. Thus, it is impractical to use the centralized DRL algorithm to make decisions. In addition, the optimal policy needs to search $8^8 = 1.68\times10^7$ possible index sets to obtain the optimal decision. Thus, it is too complicated to obtain the performance of the optimal policy. Therefore, in Fig. \ref{fig:DDRLi}, we only show the performance of the proposed distributed DRL algorithm and the random policy. From this figure, we can see that the average sum transmission rate is about $1.1$ bits/frame/Hz for the proposed distributed DRL algorithm, while the rate for the random policy is about $0.6$ bits/frame/Hz. This observation indicates that the proposed distributed DRL algorithm is effective when the number of the IoT devices and the number of the cellular users are large.

\section{Conclusions}
\label{sec:conclusions}

This paper has studied the user association problem in AmBC-based SRN using the DRL approaches. Since it is difficult to obtain the full real-time channel information, we use the historical information to infer the current information by the DRL approaches to make appropriate decisions. In particular, we propose two DRL algorithms, namely, centralized DRL and distributed DRL. The centralized DRL algorithm involves the globally available information as current state and outputs one action that involves decisions for all IoT device. While the distributed DRL algorithm uses the locally available information as current state and outputs decisions individually for each IoT device.
Finally, simulation results have demonstrated that the two DRL algorithms can perform close to the optimal policy with perfect real-time information. In addition, the centralized DRL algorithm needs less information than the distributed DRL algorithm, while the distributed DRL algorithm has the advantage of scalability, which means it can still work effectively even when the number of the IoT devices changes.


\begin{thebibliography}{10}
\providecommand{\url}[1]{#1}
\csname url@samestyle\endcsname
\providecommand{\newblock}{\relax}
\providecommand{\bibinfo}[2]{#2}
\providecommand{\BIBentrySTDinterwordspacing}{\spaceskip=0pt\relax}
\providecommand{\BIBentryALTinterwordstretchfactor}{4}
\providecommand{\BIBentryALTinterwordspacing}{\spaceskip=\fontdimen2\font plus
\BIBentryALTinterwordstretchfactor\fontdimen3\font minus
  \fontdimen4\font\relax}
\providecommand{\BIBforeignlanguage}[2]{{%
\expandafter\ifx\csname l@#1\endcsname\relax
\typeout{** WARNING: IEEEtran.bst: No hyphenation pattern has been}%
\typeout{** loaded for the language `#1'. Using the pattern for}%
\typeout{** the default language instead.}%
\else
\language=\csname l@#1\endcsname
\fi
#2}}
\providecommand{\BIBdecl}{\relax}
\BIBdecl

\bibitem{andrews2014will}
J.~G. Andrews, S.~Buzzi, W.~Choi, S.~V. Hanly, A.~Lozano, A.~C. Soong, and
  J.~C. Zhang, ``What will {5G} be?'' \emph{IEEE J. Sel. Areas Commun.},
  vol.~32, no.~6, pp. 1065--1082, Jun. 2014.

\bibitem{Wang2017A}
Y.-P.~E. {Wang}, X.~{Lin}, A.~{Adhikary}, A.~{Grovlen}, Y.~{Sui},
  Y.~{Blankenship}, J.~{Bergman}, and H.~S. {Razaghi}, ``A primer on {3GPP}
  narrowband {Internet of Things},'' \emph{IEEE Commun. Mag.}, vol.~55, no.~3,
  pp. 117--123, March 2017.

\bibitem{zhang2016spectrum}
L.~Zhang, Y.-C. Liang, and M.~Xiao, ``Spectrum sharing for {Internet of
  Things}: A survey,'' \emph{arXiv preprint arXiv:1810.04408}, 2018.

\bibitem{zhang2019backscatter}
Q.~{Zhang}, L.~{Zhang}, Y.~{Liang}, and P.~{Kam}, ``Backscatter-{NOMA}: A
  symbiotic system of cellular and {Internet-of-Things} networks,'' \emph{IEEE
  Access}, vol.~7, pp. 20\,000--20\,013, 2019.

\bibitem{long2019full}
R.~{Long}, H.~{Guo}, L.~{Zhang}, and Y.~{Liang}, ``Full-duplex backscatter
  communications in symbiotic radio systems,'' \emph{IEEE Access}, vol.~7, pp.
  21\,597--21\,608, 2019.

\bibitem{guo2019resource}
H.~{Guo}, Y.~{Liang}, R.~{Long}, S.~{Xiao}, and Q.~{Zhang}, ``Resource
  allocation for symbiotic radio system with fading channels,'' \emph{IEEE
  Access}, vol.~7, pp. 34\,333--34\,347, 2019.

\bibitem{liu2013ambient}
V.~Liu, A.~Parks, V.~Talla, S.~Gollakota, D.~Wetherall, and J.~R. Smith,
  ``Ambient backscatter: Wireless communication out of thin air,'' \emph{Proc.
  ACM SIGCOMM}, vol.~43, no.~4, pp. 39--50, Oct. 2013.

\bibitem{wang2016ambient}
G.~Wang, F.~Gao, R.~Fan, and C.~Tellambura, ``Ambient backscatter communication
  systems: Detection and performance analysis,'' \emph{IEEE Trans. Commun.},
  vol.~64, no.~11, pp. 4836--4846, Nov. 2016.

\bibitem{qian2017semi}
J.~Qian, F.~Gao, G.~Wang, S.~Jin, and H.~Zhu, ``Semi-coherent detection and
  performance analysis for ambient backscatter system,'' \emph{IEEE Trans.
  Commun.}, vol.~65, no.~12, pp. 5266--5279, Dec. 2017.

\bibitem{ZhangLiangGlobecom17}
Q.~Zhang and Y.-C. Liang, ``Signal detection for ambient backscatter
  communications using unsupervised learning,'' in \emph{IEEE GLOBECOM Workshop
  2017}, Singapore, Dec. 2017, pp. 1--6.

\bibitem{yang2018modulation}
G.~Yang, Y.-C. Liang, R.~Zhang, and Y.~Pei, ``Modulation in the air:
  Backscatter communication over ambient {OFDM} carrier,'' \emph{IEEE Trans.
  Commun.}, vol.~66, no.~3, pp. 1219--1233, Mar. 2018.

\bibitem{yang2018cooperative}
G.~Yang, Q.~Zhang, and Y.-C. Liang, ``Cooperative ambient backscatter
  communications for green {Internet-of-Things},'' \emph{IEEE Internet Things
  J.}, vol.~5, no.~2, pp. 1116--1130, Apr. 2018.

\bibitem{guo2019exploiting}
H.~Guo, Q.~Zhang, S.~Xiao, and Y.-C. Liang, ``Exploiting multiple antennas for
  cognitive ambient backscatter communication,'' \emph{IEEE Internet Things
  J.}, vol.~6, no.~1, pp. 765--775, 2019.

\bibitem{zhang2019constellation}
Q.~Zhang, H.~Guo, Y.-C. Liang, and X.~Yuan, ``Constellation learning-based
  signal detection for ambient backscatter communication systems,'' \emph{IEEE
  J. Sel. Areas Commun.}, vol.~37, no.~2, pp. 452--463, 2019.

\bibitem{kang2018riding}
X.~Kang, Y.-C. Liang, and J.~Yang, ``Riding on the primary: A new spectrum
  sharing paradigm for wireless-powered {IoT} devices,'' \emph{IEEE Trans. on
  Wireless Commun.}, vol.~17, no.~9, pp. 6335--6347, Sep. 2018.

\bibitem{luong2018applicationsDeep}
N.~C. Luong, D.~T. Hoang, S.~Gong, D.~Niyato, P.~Wang, Y.-C. Liang, and D.~I.
  Kim, ``Applications of deep reinforcement learning in communications and
  networking: A survey,'' \emph{arXiv preprint arXiv:1810.07862}, 2018.

\bibitem{anh2018deep}
T.~T. Anh, N.~C. Luong, D.~Niyato, Y.-C. Liang, and D.~I. Kim, ``Deep
  reinforcement learning for time scheduling in {RF}-powered backscatter
  cognitive radio networks,'' \emph{arXiv preprint arXiv:1810.04520}, 2018.

\bibitem{chu2018reinforcement}
M.~Chu, H.~Li, X.~Liao, and S.~Cui, ``Reinforcement learning based multi-access
  control with energy harvesting,'' in \emph{2018 IEEE GLOBECOM}.\hskip 1em
  plus 0.5em minus 0.4em\relax IEEE, 2018, pp. 1--6.

\bibitem{zhang2018deep}
L.~Zhang, J.~Tan, Y.-C. Liang, G.~Feng, and D.~Niyato, ``Deep reinforcement
  learning based modulation and coding scheme selection in cognitive
  heterogeneous networks,'' \emph{arXiv preprint arXiv:1811.02868}, 2018.

\bibitem{he2017deep}
Y.~He, Z.~Zhang, F.~R. Yu, N.~Zhao, H.~Yin, V.~C. Leung, and Y.~Zhang,
  ``Deep-reinforcement-learning-based optimization for cache-enabled
  opportunistic interference alignment wireless networks,'' \emph{IEEE Trans.
  Veh. Technol.}, vol.~66, no.~11, pp. 10\,433--10\,445, 2017.

\bibitem{wang2018handover}
Z.~Wang, L.~Li, Y.~Xu, H.~Tian, and S.~Cui, ``Handover control in wireless
  systems via asynchronous multiuser deep reinforcement learning,'' \emph{IEEE
  Internet Things J.}, vol.~5, no.~6, pp. 4296--4307, 2018.

\bibitem{yu2018deep}
Y.~Yu, T.~Wang, and S.~C. Liew, ``Deep-reinforcement learning multiple access
  for heterogeneous wireless networks,'' in \emph{2018 IEEE ICC}.\hskip 1em
  plus 0.5em minus 0.4em\relax IEEE, 2018, pp. 1--7.

\bibitem{zhao2018deep}
N.~Zhao, Y.-C. Liang, D.~Niyato, Y.~Pei, and Y.~Jiang, ``Deep reinforcement
  learning for user association and resource allocation in heterogeneous
  networks,'' in \emph{2018 IEEE GLOBECOM}.\hskip 1em plus 0.5em minus
  0.4em\relax IEEE, 2018, pp. 1--6.

\bibitem{sun2018smart}
Y.~Sun, G.~Feng, S.~Qin, Y.-C. Liang, and T.-S.~P. Yum, ``The {SMART} handoff
  policy for millimeter wave heterogeneous cellular networks,'' \emph{IEEE
  Trans. Mobile Comput.}, vol.~17, no.~6, pp. 1456--1468, 2018.

\bibitem{nasir2018deep}
Y.~S. Nasir and D.~Guo, ``Deep reinforcement learning for distributed dynamic
  power allocation in wireless networks,'' \emph{arXiv preprint
  arXiv:1808.00490}, 2018.

\bibitem{liang2017spectrum}
L.~{Liang}, J.~{Kim}, S.~C. {Jha}, K.~{Sivanesan}, and G.~Y. {Li}, ``Spectrum
  and power allocation for vehicular communications with delayed csi
  feedback,'' \emph{IEEE Wireless Commun. Lett.}, vol.~6, no.~4, pp. 458--461,
  Aug 2017.

\bibitem{liang1999downlink}
Y.-C. {Liang}, P.~S.~F. {Chin}, and K.~J.~R. {Liu}, ``Downlink beamforming for
  {DS-CDMA} mobile radio with multimedia services,'' \emph{IEEE Trans.
  Commun.}, vol.~49, no.~7, pp. 1288--1298, Jul. 2001.

\bibitem{mnih2015human}
V.~Mnih, K.~Kavukcuoglu, D.~Silver, A.~A. Rusu, J.~Veness, M.~G. Bellemare,
  A.~Graves, M.~Riedmiller, A.~K. Fidjeland, G.~Ostrovski \emph{et~al.},
  ``Human-level control through deep reinforcement learning,'' \emph{Nature},
  vol. 518, no. 7540, p. 529, 2015.

\bibitem{kaelbling1996reinforcement}
L.~P. Kaelbling, M.~L. Littman, and A.~W. Moore, ``Reinforcement learning: A
  survey,'' \emph{J. Artificial Intell. Research}, vol.~4, pp. 237--285, 1996.

\bibitem{sutton2018reinforcement}
R.~S. Sutton and A.~G. Barto, \emph{Reinforcement learning: An
  introduction}.\hskip 1em plus 0.5em minus 0.4em\relax MIT press, 2018.

\bibitem{calabrese2016learning}
F.~D. {Calabrese}, L.~{Wang}, E.~{Ghadimi}, G.~{Peters}, L.~{Hanzo}, and
  P.~{Soldati}, ``Learning radio resource management in rans: Framework,
  opportunities, and challenges,'' \emph{IEEE Commun. Mag.}, vol.~56, no.~9,
  pp. 138--145, Sep. 2018.

\end{thebibliography}

\end{document}